\documentclass[twocolumn,amssymb,nobibnotes,aps,showpacs,prl]{revtex4}
  \usepackage[dvips]{color}
  \usepackage{newlfont}
  \usepackage{graphicx} 
  \usepackage{dcolumn}
  \usepackage{amssymb}
  \usepackage{amsfonts}
  \usepackage{amsmath}
  \usepackage[latin1]{inputenc}
  \usepackage{float}
\newcommand{\br}{\begin{eqnarray}}
\newcommand{\er}{\end{eqnarray}}

\newcommand{\bra}{\langle}
\newcommand{\ket}{\rangle}

\def\rg{\rangle}
\def\lg{\langle}

\begin{document}
\title{ Generalized Purity and Quantum Phase Transition for Bose-Einstein condensates in a Symmetric Double Well }
\author{Thiago F. Viscondi}
\affiliation{Instituto de F\'{i}sica ``Gleb Wataghin'',  Universidade
Estadual de Campinas, 13083-970, Campinas - SP, Brazil}
\author{K. Furuya}
\affiliation{Instituto de F\'{i}sica ``Gleb Wataghin'',  Universidade
Estadual de Campinas, 13083-970, Campinas - SP, Brazil}
\author{M.C. de Oliveira}
\affiliation{Instituto de F\'{i}sica ``Gleb Wataghin'',  Universidade
Estadual de Campinas, 13083-970, Campinas - SP, Brazil}

\begin{abstract}
The generalized purity is employed for investigating the process of coherence loss and delocalization of the Q-function in the Bloch sphere of a two-mode Bose-Einstein condensate in a symmetrical double well with cross-collision. Quantum phase transition of the model is signaled by the generalized purity as a function of an appropriate parameter of the Hamiltonian and the number of particles ($N$). A power law dependence of the critical parameter with $N$ is derived.

\end{abstract}

\maketitle


Recently it has been pointed out that any bipartite and multipartite entanglement measure can 
signalize the presence of a quantum phase transition (QPT) \cite{wu04,tro1, amico08} in many particle systems.
 Related to that, a subsystem-independent generalization of entanglement has been introduced
based on coherent states and convex sets characterizing
the unentangled pure states as coherent states of a chosen Lie algebra  \cite{BKOV03}.
Such a notion of entanglement defined relative 
to a distinguished subspace of observables is pointed as particularly
useful for classifying multipartite entanglement and thus for QPT characterization. 
 The  generalized purity (GP) of the state relative 
to a certain distinguished subset of observables forming a local Lie 
algebra is directly related to the Meyer-Wallach  measure  
\cite{MW02}, and whenever a specific subsystem can be associated to 
the subset of observables the usual entanglement notion is recovered. 
On the other hand it has been demonstrated that the model describing a two mode Bose-Einstein condensate (BEC) in a symmetric double well (BECSDW) \cite{milburn}, with cross-collisional terms \cite{bruno,thiago} presents interesting dynamical regimes: The { macroscopic self-trapping (MST)} and Josephson oscillations (JO) of population \cite{milburn,smerzi,thiago}, both experimentally observed \cite{albiez}. 
Recently a discussion of the transition from one regime to the other
has been presented from the point of view of critical phenomena, as
a continuous quantum phase transition problem in terms of the
usual subsystem entropy \cite{HMM03,fu06,vidal04}. 

In this paper we apply the concept of GP in a BECSDW.
Our purpose is twofold:
{\em (i)} We employ the GP relative to a chosen set of 
observables to quantify the quality of the semiclassical
approach used in most of the treatments of BECSDW; {\em (ii)} We
characterize the quantum critical phenomena occurring in this model    
 with the subsystem independent measure of quantum correlations \cite{BKOV03}.
The two mode approximated BECSDW 
has been well studied in the literature (see e.g. \cite{milburn,smerzi} )
as a model presenting a nonlinear self-trapping phenomena. 
More recently, the importance of the cross-collisional terms for 
large number of particles in the condensate ($N\gg  1$) has 
been noticed \cite{bruno} and explored semiclassically with a
time dependent variational principle (TDVP) based on coherent states 
\cite{thiago}. 
For a fixed number of condensed particles $N$, in order 
to explore the  natural group structure of the model, we conveniently 
adopt the  Schwinger's pseudo-spin operators  
 defined in terms of the creation and 
annihilation boson operators $d_{\pm}^{\dag},d_{\pm}$ on the approximated
localized states  $|u_{\pm}\rangle$ \cite{milburn,thiago}: $J_{x}\equiv ({d_{-}^{\dag}d_{-}-d_{+}^{\dag}d_{+}})/{2}$,
$J_{y}\equiv i({d_{-}^{\dag}d_{+}-d_{+}^{\dag}d_{-}})/{2}$, and
 $J_{z}\equiv ({d_{+}^{\dag}d_{-}+d_{-}^{\dag}d_{+}})/{2}$, 
 where $J=N/2$. In that form the two-mode BEC Hamiltonian writes 
as 
\br
\hat{H} = 2\left[2\Lambda(N-1)+\frac{\Omega}{2}\right]J_{z}+2(\kappa-\eta)J_{x}^{2}+4\eta
J_{z}^2.
\label{hamil2}
\er
where $\Omega$ is the tunneling parameter, $\kappa$ is the self-collision 
parameter of the condensate which is much larger than the so called 
cross-collision terms $\eta=\kappa \epsilon^2,\, \Lambda=\kappa \epsilon^{\frac{3}{2}}$, with 
 $\epsilon = \lg u_+|u_-\rg$. 
 $\Omega^{'}\equiv 2[2\Lambda(N-1)+ \Omega/2]$ is an effective tunneling parameter dependent on $N$. 
 The natural associated 
algebra of the model is $su(2)$.  In this form the Hamiltonian (\ref{hamil2}) is a realization of the Lipkin-Meshkov-Glick model \cite{LMG}, whose ground state entanglement has been investigated recently \cite{vidal04} in the $\eta=0$ limit .


By means
of a semiclassical method exploring $SU(2)$ coherent states we have shown   that for a 
sufficiently large $N$ even a small amount of cross-collisional rate can 
change the dynamical regime of oscillations of the condensate from 
MST to JO \cite{thiago}. 
 The quantum and semiclassical MST and JO dynamics of the BECSDW are known 
to be qualitatively  very similar \cite{milburn,thiago} for 
large number of particles ($N \gg 1$), except for the presence of collapses and 
revivals in the quantum time evolutions for the mean values of 
$\langle J_x \rangle (t)$.
 This breaking of 
quantum-classical correspondence  is due to  {\em decoherence} 
\footnote{ Here {\em decoherence} is employed as 
{ being far from the classical situation}, not to be confused with the more frequently 
used sense where the classical character is induced by a coupling with the 
environment.
} 
 introduced by quantum fluctuations, which drives the 
state of the system away from a coherent state. 
 When we treat the semiclassical dynamics employing  a TDVP \cite{TDVP}, we restrict the evolution of the 
state to a  nonlinear subspace constituted only by the coherent states of $SU(2)$ \cite{thiago}. Such an evolution is exact for an initially coherent state 
{\em only} in the macroscopic limit of $N\rightarrow\infty$ 
\footnote{This is also true when the Hamiltonian is linear in the generators of 
the dynamical group. Namely this would be the present case only in the limit
of non-interacting particles ($\kappa=0=\eta$).}. 
 Also, in the TDVP approach 
we force the system wave function to be 
always localized in the phase space. 
 From this point of view, the coherent states are the closest to the 
classical ones, which are points in the classical phase space, and converges to 
them in the limit $N\rightarrow\infty$. 
The { delocalization of the 
wave function (DWF)} which goes along with the decoherence, is the responsible 
for the quantitative disagreement between the two 
dynamics \cite{thiago}, when collapses and 
revivals of expectation values of relevant observables happens with 
the DWF and consequent self-interferences occasioning the appearance of 
superposition states. 

For a more complete analysis of the decoherence or of the {\em quality
of the semiclassical approximation}, it becomes fundamental to have a
good quantitative measure of the `distance' of a given state to the subspace
of the coherent states that give birth to the classical phase space. 
 When  the complete dynamics of the system is
restricted to a space that carries an irreducible representation of $SU(2)$
(here this condition is satisfied due to the particle number 
conservation), there is a  simple measure 
 called {  GP of the algebra}  $\mathcal{P}_{su(2)}$ 
\cite{BKOV03,klyachko}, defined for  $su(2)$ as
\begin{equation}
\mathcal{P}_{su(2)}(|\psi\ket)=\frac{1}{J^{2}}\sum_{k=x,y,z} \bra\psi|J_{k}|\psi\ket^{2}.
\label{pdup59}
\end{equation} It is 
 a good measure of decoherence, among 
other reasons, because it is invariant under a transformation of the group 
$SU(2)$ on the state $|\psi\ket$: $ \mathcal{P}_{su(2)}(|\psi\ket)=\mathcal{P}_{su(2)}(U|\psi\ket),\;\forall\;\;U\in SU(2)$. Thus, all the states connected by a  $SU(2)$ transformation possesses
the same purity. However, the most interesting property of 
$\mathcal{P}_{su(2)}$ -- concerning the purpose of quantitatively compare
the correspondence between the semiclassical and the exact quantum 
dynamics -- is the fact that this measure has its maximum value at one,
{\em if and only if}, the state is the coherent state closest to the 
classical state:
$ %
|\theta,\phi\ket=R(\theta,\phi)|J,-J\ket 
=e^{-i\theta(J_{x}\sin\phi-J_{y}\cos\phi)}|J,-J\ket
%
$.
As soon as such a state moves away from the coherence, becoming delocalized
in the phase space, the  GP  decreases monotonically to 
zero. 
 Remark that the GP only has such properties clearly defined for pure states and, only in this case, it is a measure of existing quantum correlations of the state on the classical phase
space. 
\begin{figure}[ht]
    \centering
    \includegraphics[height=6cm]{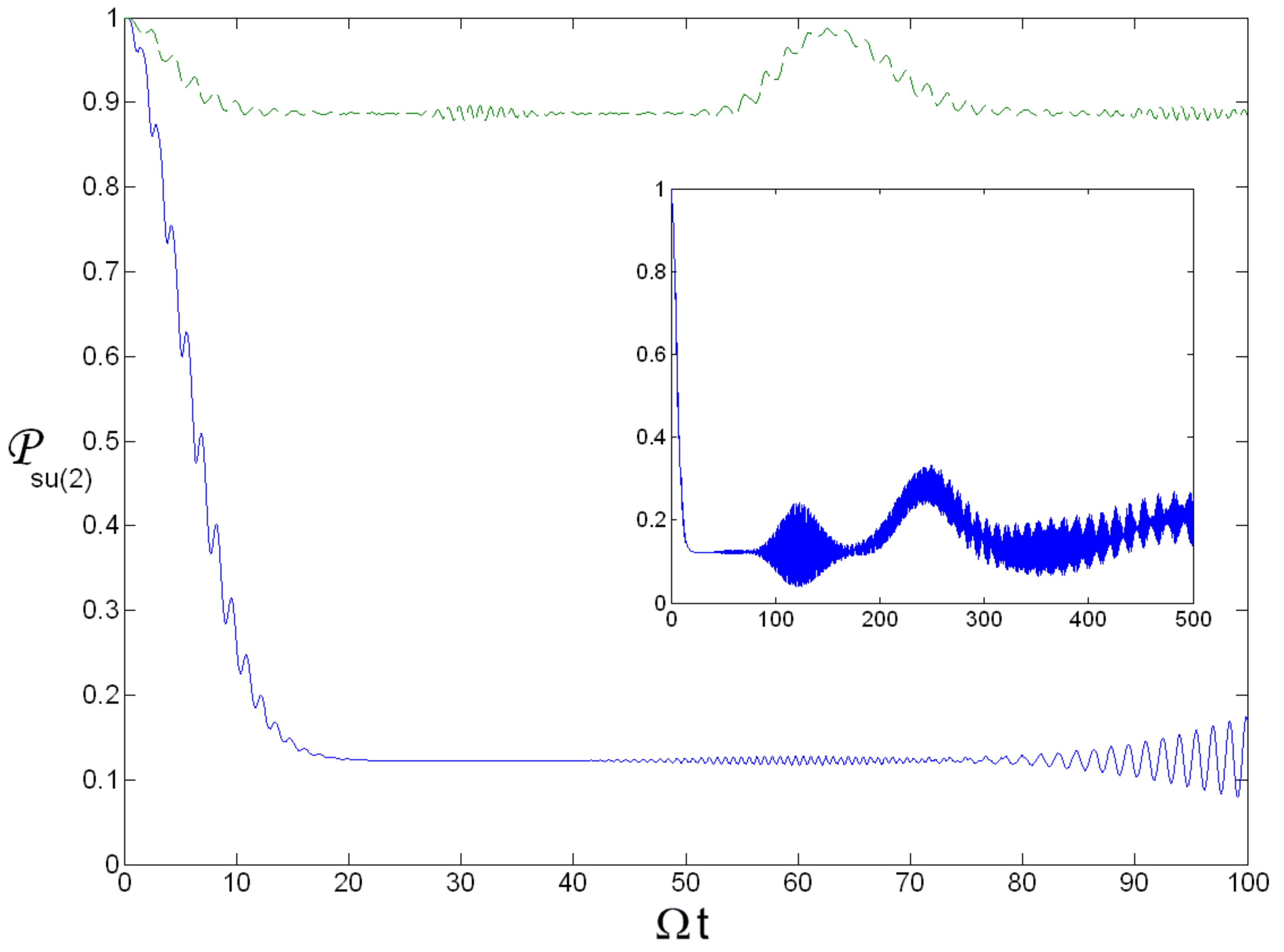}
    \caption{(color online) Time evolution of the  GP $\mathcal{P}_{su(2)}$ 
for MSP and JO dynamical regimes of the two mode BEC model.
The dashed curve represents the MST regime with
    $\kappa=\frac{2\Omega}{N}$ and $\eta=\frac{\kappa}{100}$, whereas
the solid curve represents the regime of JO with
    $\kappa=\frac{2\Omega}{N}$ e $\eta=\frac{\kappa}{10}$.
For both cases the initial state of the system is the CS
    $|J,J\ket_{x}\equiv|\theta=\frac{\pi}{2},\phi=0\ket$,
where $N=100$ particles are initially in the same well.
}
\label{fig1}
\end{figure}

We choose  $|J,J\ket_{x}\equiv|\theta=\frac{\pi}{2},\phi=0\ket$
where $N = 100$ particles are in the same well, $\Omega = 1$ and 
$\kappa=\frac{2\Omega}{N}$. For this set of parameters it is known
that choosing $\eta=\frac{\kappa}{100}$ the state is in the self-trapping region 
of the phase space, whereas for $\eta=\frac{\kappa}{10}$ it is outside
the self-trapping region and thus in the JO regime  \cite{thiago}.
 In Fig.(\ref{fig1}) we plot the GP, where the MST regime is in dashed line, 
and solid one for JO regime.
In the MST regime the GP quickly drops down from 1 stabilizing at $0.9$ for $\Omega t\approx10$, indicating that the dynamics takes the state away from the subspace of CS. This 
 plateau of purity coincides with the collapse region of population dynamics \cite{thiago}. Note that close to  $\Omega t=30$ the purity shows small oscillations and in the region close to $\Omega t=60$ the purity increases again until it practically recovers the value 1. At this instant the re-coherence of the state happens, being responsible for the revival of the oscillations of the population dynamics \cite{thiago}. 
 The agreement between the quantum and semiclassical results is due to the oscillation of the mean value of the generator $J_{x}$ in the MST, since the purity depends on the normalized square of such mean value. The oscillations of $\bra J_{x}\ket$ around a non-zero value close to  $J$ keeps also the value of $\mathcal{P}_{su(2)}$ close to its maximum possible value. In the JO regime %
 for  $\kappa=\frac{2\Omega}{N}$ and $\eta=\frac{\kappa}{10}$, the purity decays rapidly with the time, but in this case the decoherence is much stronger, and the purity reaches much lower values, not recovering to 1. Hence, this regime presents lower quantitative agreement between the quantum and classical evolutions, when compared to the MST regime. In this dynamical regime, the system does not recover high values of coherence. At a time close to $\Omega t=250$ the state reaches its maximum re-coherence, but the GP value is still lower than $0.4$. This result is expected because the classical orbit delocalizes much more on the Bloch sphere for this regime, and this entails a correspondingly large delocalization of the semiclassical Q-function on the sphere, 
$Q(\theta,\phi)=\bra\theta,\phi|\rho|\theta,\phi\ket$, with $ \rho=|\psi\ket\bra\psi|$,
%
 during its evolution \cite{thiago,trimborn}. 
 The larger 
the region traveled by the trajectory in the phase space, larger the 
broadening of the distribution and smaller the coherence left on the state. 
Therefore, when the Hamiltonian has nonlinear terms in the generators of the
dynamical group, the semi-classical approximation has better
quantitative accordance with the exact quantum results (for finite $N$) 
 when the classical orbits sweep smaller `volume' in the phase
space. Since BECSDW model is integrable, we cannot analyze the 
decoherence due to chaotic trajectories. However, our results indicate 
that the semiclassical method has lower validity
for this type of trajectory that is less localized in the phase space.

Now we can take advantage of such qualities of the GP to characterize  the { quantum 
phase transition} (QPT) \cite{amico08}. The QPT is connected to a non-analyticity of the 
energy of the fundamental state of the system, when it is taken as a function 
of some real continuum parameter of the Hamiltonian \cite{wu04} at zero 
temperature, in the thermodynamic limit  $N\rightarrow\infty$.
Generally speaking, the energy of the ground state in a finite system is 
an analytic function of any parameter of the Hamiltonian and only shows
non-analyticity when $N\rightarrow\infty$, which then corresponds to an 
avoided level crossing. However, even when we are not allowed to take 
effectively such a limit, we still can observe the scaling of the 
properties of the system for increasing $N$ and infer about the occurrence 
of the QPT in the thermodynamic limit.

The BECSDW 
suffers a sudden change in its dynamics when $\kappa_{c}=\frac{\Omega}{2N}$ in the limit of no cross-collision terms
and $N\gg1$. With $\Omega,\kappa>0$, such a
transition of regime (JO to MST) does not occur at the ground state, but
at the largest energy state for the value of parameter 
corresponding to the bifurcation in the classical phase space, which causes
the appearance of a separatrix of motion.
Strictly speaking, 
the QPT is only characterized in the limit $N\rightarrow\infty$, thus our 
transition of dynamical regime (even if it occurred at the ground state) 
would only be considered as a {\em bona fide } QPT in the exact classical limit.  
\begin{figure}[ht]\vspace{-2.5cm}
   \centering
   \includegraphics[height=11cm]{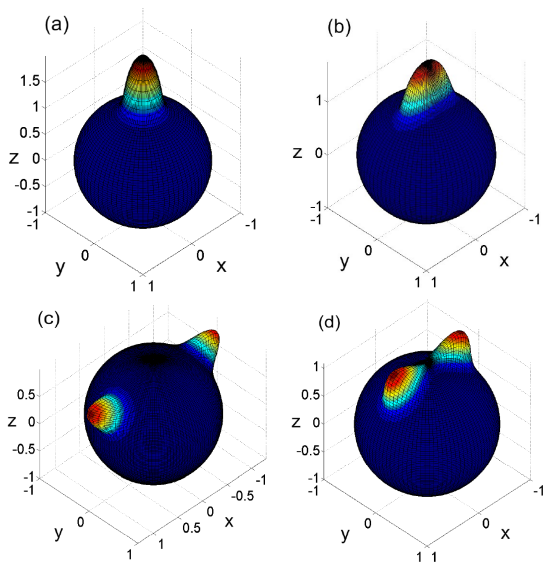}\vspace{-1cm}
   \caption{(Color online) Q-function for the largest energy state of 
the $\hat{H}$ spectrum for several value of parameters $\kappa$ and 
$\eta$. Considering the number of particles $N=100$ and $\Omega=1$, we 
have the following values for the collision rates: 
$(a)$ $\kappa=\eta=0$; $(b)$ $\kappa=\frac{\Omega}{2N}$,$\eta=0$; 
$(c)$ $\kappa=\frac{\Omega}{N}$,     $\eta=0$ and
$(d)$ $\kappa=\frac{2\Omega}{N}$,    $\eta=\frac{\kappa}{10}$.}
    \label{fig2}
\end{figure}
In Fig.(\ref{fig2}), we have the Q-functions for the eigenstate of largest 
energy of the spectra of $\hat{H}$ for several values of parameters $\kappa$ 
and $\eta$. Fig.(\ref{fig2})$a$  just shows the coherent state
$|\theta=\pi,\phi\ket=|J,J\ket_{z}$, which is the maximum energy state of the 
non-interacting case  $\kappa=\eta=0$. In the absence of collisions, the
most energetic eigenstate corresponds to the most localized state in the 
phase space, such that $\mathcal{P}_{su(2)}=1$. Increasing $\kappa$, but still not considering the cross-collision
terms, the Q-function broadens along the $x$-axis, and consequently we expect the decreasing of the GP. At $\kappa=\kappa_{c}$, as 
shown in Fig.(\ref{fig2})$b$, the state is greatly broadened, but still does not 
show a bifurcation; namely, the formation of two maxima in its distribution.
This behavior is expected, since for finite $N$, the quantum transition 
parameter $\kappa_{c}^{q}(N)$ is slightly different from the value of transition
$\kappa_{c}$ of the classical limit. But for increased $\kappa$, such as in  Fig.(\ref{fig2})$c$, for $\kappa=\frac{\Omega}{N}$  we see two maxima far apart along the $x$-axis as a signature of the bifurcation.
The increase of the cross-collision has the opposite effect. For $\kappa=\frac{2\Omega}{N}$, $\eta=\frac{\kappa}{10}$ the two peaks of the Q-function become closer, as in Fig.(\ref{fig2})$d$.
\begin{figure}[ht]
    \centering
    \includegraphics[height=5cm]{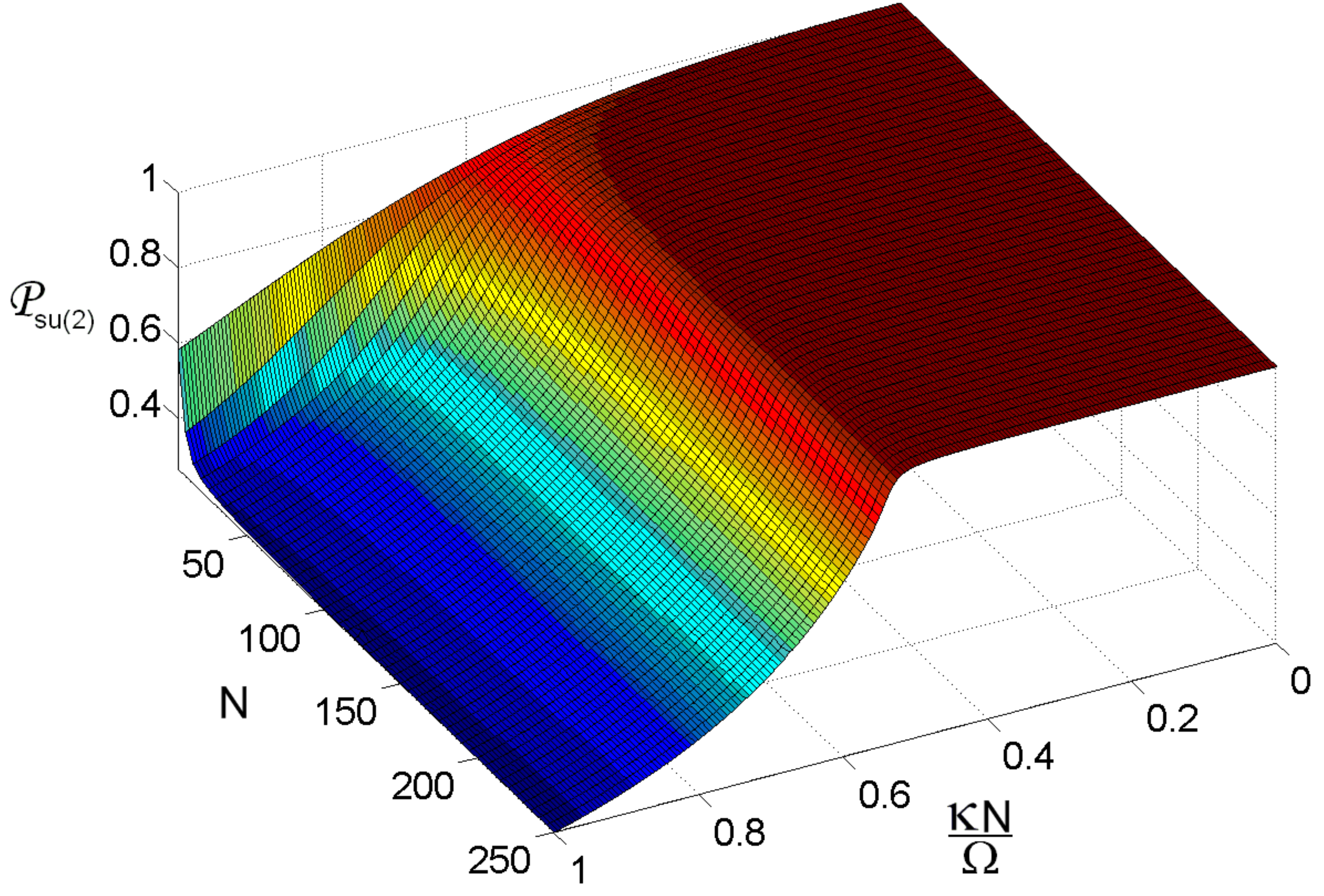}
    \caption{(Color online) GP of $su(2)$ for the largest energy eigenstate
as a function of normalized self-collision parameter and the number of 
particles, with cross-collision rate $\eta=0$.
}
    \label{fig5}
\end{figure}

Our results for the phase space distribution of the maximum energy state
are confirmed as we analyze the behavior of the  GP as a
function of the self-collision parameter and the number of particles, 
as shown in Fig.(\ref{fig5}), neglecting the cross-collisions. The GP
initially decreases slowly with $\frac{\kappa N}{\Omega}$, independent
of the value of $N$, corresponding to the region where the distributions
 broadens along the $x$-axis. However, 
close to  $\frac{\kappa N}{\Omega}=\frac{1}{2}$, the GP begins to decrease
quickly and, although smoothly, the lowering of its value is more and more
steep as we increase $N$. This behavior of $\mathcal{P}_{su(2)}$ suggests
us a strong dependence between the derivative of the GP with respect to
 $\frac{\kappa N}{\Omega}$ and the number of particles.
\begin{figure}[ht]
    \centering
    \includegraphics[height=5cm]{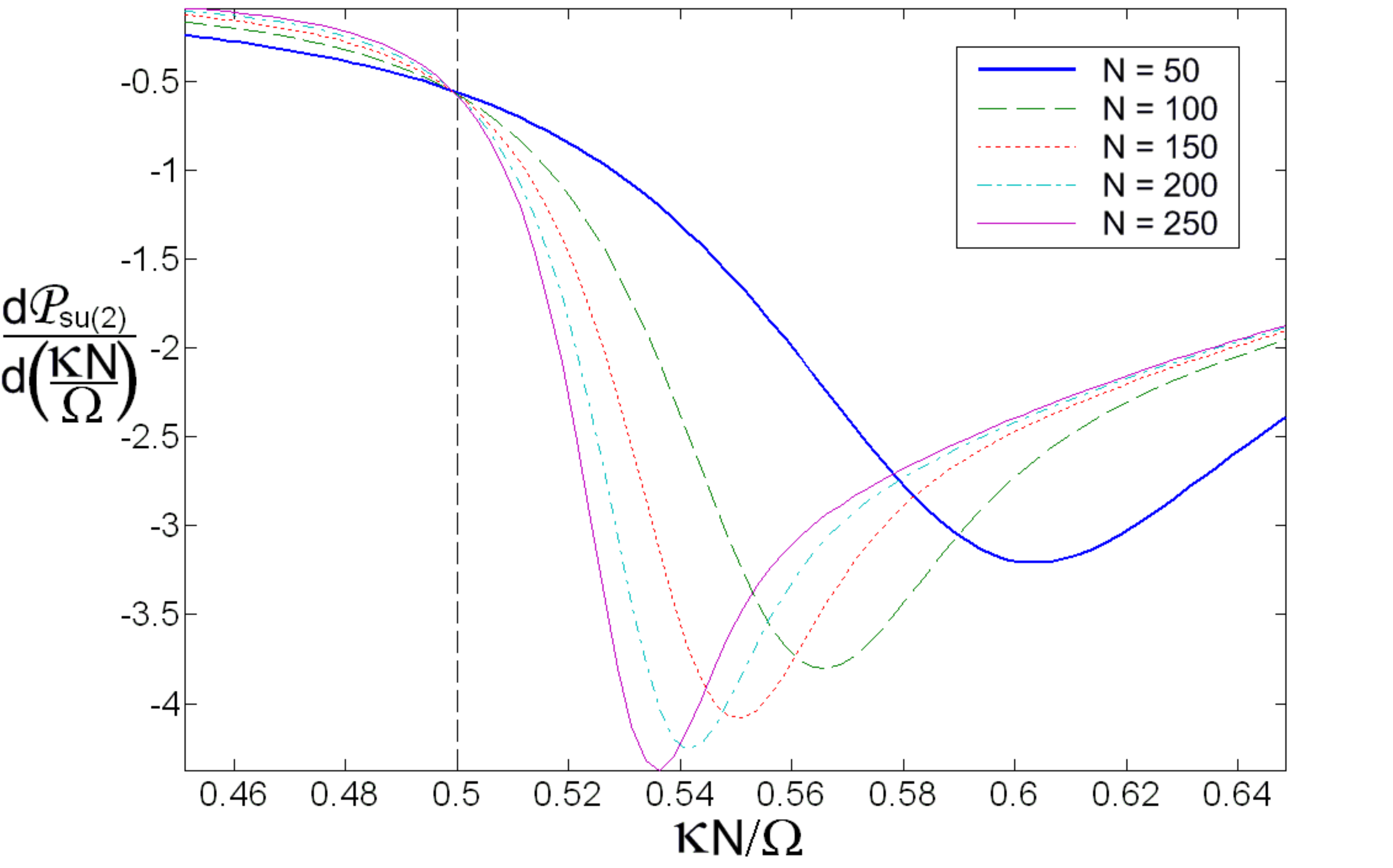}
    \caption{(Color online) Derivative of the  GP of the eigenstate of largest
energy eigenvalue with respect to $\frac{\kappa N}{\Omega}$ for various values of the total number of particles
$N$. The dashed vertical line represents the position at the value of the 
classical transition point  $\frac{\kappa_{c}N}{\Omega}=\frac{1}{2}$. 
All the curves for finite systems cross exactly at this critical
value.
}
    \label{fig6}
\end{figure}
In Fig.(\ref{fig6}) we show the derivative of the GP with respect to
the normalized self-collision parameter $\frac{\kappa N}{\Omega}$  for
various values of $N$. For an increasing number of particles, we see
the minimum value of the derivative of  $\mathcal{P}_{su(2)}$ to move to 
the left side, closer to the classical critical value ($0.5$), and also
the minimum becomes more pronounced. We define the value of $\frac{\kappa N}{\Omega}$ at
the minimum of the derivative of GP as the critical value of the quantum
dynamical transition $\kappa_{c}^{q}(N)$. 
It is already clear that the value of $\kappa_{c}^{q}(N)$ is brought closer
to the classical transition value $\kappa_{c}=\frac{\Omega}{2N}$ for increasing
 $N$, but we still need to characterize how this approximation happens. 
The values of 
$\ln\left[\frac{N(\kappa_{c}^{q}-\kappa_{c})}{\Omega}\right]$ from the curves in Fig. (\ref{fig6}) suggest a power law between  $(\kappa_{c}^{q}-\kappa_{c})$ and $N$. A linear 
interpolation of the data points gives %
$\kappa_{c}^{q}-\kappa_{c}=\frac{\Omega}{N}e^{0,31\pm                          
0,05}N^{-0,657\pm0,009}\propto N^{-1,657\pm0,009}$.  
%
It is evident that 
$\kappa_{c}^{q}\rightarrow\kappa_{c}$  as  $N\rightarrow\infty$.


In conclusion we considered the GP to analyze the dynamics 
of coherence loss of initially coherent state for the BECSDW model.
In the MST regime 
 the GP remains high without significant decoherence. 
For the JO regime, on the other hand, 
we have seen that the first decay of the GP was in a similar 
time scale to the MST one, being the decoherence 
much more significant. In the JO regime no considerable re-coherence can 
be achieved 
 and the GP correspondingly has a much lower value than in the previous 
regime at this time. Since the coherent state represents the closest 
to the classical state, the value of GP enabled us to estimate the quality
of the semiclassical approximation at each time in both regimes. Moreover
we have employed  the GP as a tool for 
characterizing a QPT in the same model. We have
shown for finite number of particles ($N < \infty$) the bifurcation of the
Q-function of the largest energy state as the self-collision 
parameter $\kappa$ becomes larger than a critical value $\kappa_c^q(N)$.
Also, we have shown the suppression of cross-collisions
between the particles in different wells. By increasing the 
number of particles $N$, the GP has shown a more and more steeper behavior
near the critical value of 
$\frac{\kappa N}{\Omega}$; moreover, its value has tended
 to the known 
classical value $\frac{\kappa_c N}{\Omega}=\frac{1}{2}$ as 
$N \rightarrow \infty$. Finally we have shown that $\kappa_c^q-\kappa_c$
is consistent with a power law in $N$.
Therefore, the GP for this model is an excellent measure 
for both to indicate the dynamical loss of coherence of initially 
coherent state and to indicate the QPT. 
When a state suffers fundamental changes resulting from
the QPT, its GP must follow its behavior, because it has all the information 
about its coherence and the degree of localization over the phase space.

\begin{acknowledgments}
{We acknowledge J. Vidal for bringing aspects of the LMG model to our knowledge. Our work is supported by FAPESP, and CNPq.}
\end{acknowledgments}

\end{document}